\documentclass[runningheads]{llncs}
\usepackage{graphicx}

\usepackage{multirow}
\usepackage{amsmath}
\usepackage{amssymb}
\usepackage{verbatim} 

\usepackage{siunitx}
\usepackage[toc]{glossaries}

\newacronym{cnn}{CNN}{convolutional neural networks}
\newacronym{auc}{AUC}{area under the curve}
\newacronym{mihc}{mIHC}{multiplex immunohistochemistry}

\newacronym{roc}{ROC}{receiver characteristic curve}

\newacronym{sgd}{SGD}{stochastic Gradient Decent}
\newacronym{g}{G}{Ground truth}
\newacronym{relu}{ReLU}{rectified linear unit}

\newacronym{concoddenet}{ConCORDe-Net}{Cell COunt RegularizeD Convolutional neural Network}

\begin{document}
\title{ConCORDe-Net: Cell Count Regularized  Convolutional Neural Network  for Cell Detection in Multiplex Immunohistochemistry Images} % \thanks{Supported by xx.}
\titlerunning{Cell detection in Multiplex Immunohistochemistry Images}

\author{Yeman Brhane Hagos\inst{1} %\orcidID{0000-0002-0357-6297} % index{Hagos, Yeman Brhane}
\and
Priya Lakshmi Narayanan\inst{1} % index{Narayanan, Priya Lakshmi}
\and  Ayse U. Akarca\inst{2} % index{Akarca, Ayse U.}
\and \\ Teresa Marafioti\inst{2} \and % index{Marafioti, Teresa}
Yinyin Yuan\inst{1}} % index{Yuan, Yinyin}
\authorrunning{Y.B. Hagos et al.}
\institute{ Division of Molecular Pathology, The Institute of Cancer Research, London, UK \and
Department of Cellular Pathology, University College London, London, UK 
}

\maketitle              

\begin{abstract}
In digital pathology, cell detection and classification are often prerequisites to quantify cell abundance and explore tissue spatial heterogeneity. However, these tasks are particularly challenging for \gls{mihc} images due to high levels of variability in staining, expression intensity, and inherent noise as a result of preprocessing artefacts. We proposed a deep learning method to detect and classify cells in \gls{mihc} whole-tumor slide images of breast cancer. Inspired by inception-v3, we developed \gls{concoddenet} which integrates conventional dice overlap and a new cell count loss function for optimizing cell detection, followed by a multi-stage convolutional neural network for cell classification. In total, 20447 cells, belonging to five cell classes were annotated by experts from $175$ patches extracted from 6 whole-tumor \gls{mihc} images. These patches were randomly split into training, validation and testing sets. Using \gls{concoddenet}, we obtained a cell detection F1 score of $0.873$, which is the best score compared to three state of the art methods. In particular, ConCORDe-Net excels at detecting closely located and weakly stained cells compared to other methods. Incorporating cell count loss in the objective function regularizes the network to learn weak gradient boundaries and separate weakly stained cells from background artefacts. Moreover, cell classification accuracy of $96.5\%$ was achieved. These results support that incorporating problem specific knowledge such as cell count into deep learning based cell detection architectures improves robustness of the algorithm.
\and Deep learning\and Breast cancer
\end{abstract}
\section{Introduction}
Cell detection and classification are often the first key steps in a wide range of histology image analysis tasks, such as investigating the interplay of the tumor and immune cells \cite{Yuan2016}. Multiplex Immunohistochemistry (mIHC) is a multi-parametric protocol that allows simultaneous examination of expression of multiple markers in a single section \cite{Blom2017,Kalra2017}. Combined with robust cell detection and classification techniques, mIHC has the potential to allow detailed  investigation of cells spatial interaction and signalling for the study of tumor heterogeneity \cite{Blom2017}. 

The field of digital pathology has recently witnessed a surge of interest in the application of deep learning for cell classification \cite{Sirinukunwattana2016a}, cell detection \cite{Raza2018,Yang2018}, and cell counting\cite{Xie2015,Rad2018,Cohen2018,Xue2016}. However, automated cell detection and classification remain challenging due to variation in slide preparation and cell morphological  diversity in shape and size. For example, closely located cells with weak boundaries are often difficult to discern \cite{Raza2018,Yang2018,Xie2015,Rad2018}. Moreover, often a parameter such as a kernel size needed to be fixed \cite{Raza2018}, which cannot cater for cells with a range of size and shape. Furthermore, the need to differentiate cells with a subtle difference in marker expression intensity, as exemplified in Fig. \ref{fig:detectionpipeline}a, adds another layer of complexity in mIHC image analysis.

In this paper, to address the above stated challenges, we developed a new cell detection method followed by multi-stage CNN to analyse mIHC images of breast cancer. Our work has the following main contributions: 1) We developed Cell Count RegularizeD Convolutional neural Network (ConCORDe-Net) inspired by inception-v3 which incorporates cell counter and designed for cell detection without the need of pre-specifying parameters such as cell size. 2) The parameters of ConCORDe-Net were optimized using an objective function that combines conventional Dice overlap and a new cell count loss function which regularizes the network parameters to detect closely located cells. 3) Our quantitative experiments support that ConCORDe-Net outperformed the state of the art methods at detecting closely located as well as weakly stained cells.
\section{Materials} \label{sec:material}
The dataset used in this paper were mIHC whole-tumor slide images from patients with breast cancer, and the images were scanned at 40X resolution. A total of $175$ regions/patches were annotated from different parts of 6 whole tumor images by experts. The patches were extracted from different regions of the slides to incorporate the variation in the data. The patches were then randomly split into training ($120$), validation ($28$), and testing ($27$).  Inside these patches $20477$ cells were annotated and these belonged to five different types of cells as depicted in Table \ref{tab:classificationdata}. Illustrative example of patches are shown in Fig. \ref{fig:detectionpipeline}a. The distribution of the data for each cell is presented in Table \ref{tab:classificationdata}.
\begin{table*}
	\centering
	\caption{Distribution of dataset}\label{tab:classificationdata}
	\begin{tabular}{|l|c|c|c|}
	\hline
	 Cell type         &  Training     & Validation & Test\\
	\hline
	CD8                  &   2971  &   653  &    624     \\
	GAL8+ pSTAT-           &  4118   &   881  &    903    \\
	GAL8+ pSTAT+ strong     &   919  &   183  &      200  \\
	GAL8+ pSTAT+ moderate    &   1558  &  295   &    279    \\
	GAL8+ pSTAT+ weak        &   4770  &    1038 &   1102 \\
	\hline
	\end{tabular}
\end{table*}
\section{Methodology}
\subsection{Dot Annotation to Cell Pseudo-segmentation}\label{sec:annotation}
The reference ground truth obtained was a dot annotation at the center of a cell instead of cell spatial extent segmentation which is generally tedious task. However, to train the proposed cell detection pipeline, cells mask ($G$) and the number of cells ($C_t$) were needed as a target. $C_t$ is simply the number of annotated cells in the input patch. Cell pseudo-segmentation was generated from dot annotation using Equation (\ref{eq:annotation}).
\begin{equation}\label{eq:annotation}
	G(i, j) = \begin{cases} 
	1 & \text{if } d < r \\
	0       & \text{} otherwise
	\end{cases}
\end{equation}
where $G (i,j)$ is pixel intensity value at $(i,j)$ of pseudo-segmentation image ($G$), $\textit{d}$ is an Euclidean distance between pixel location $(i,j)$ and any of cell dot annotations, and $\textit{r}$ is threshold distance. $\textit{r}$ was empirically set to 4 pixels to guarantee pseudo-segmentation of cells do not touch each other.
\subsection{Cell Counter}\label{sec:cellcount}
\begin{figure}
	\centering
	\begin{minipage}[b]{1.0\linewidth}
		\includegraphics[width=\textwidth]{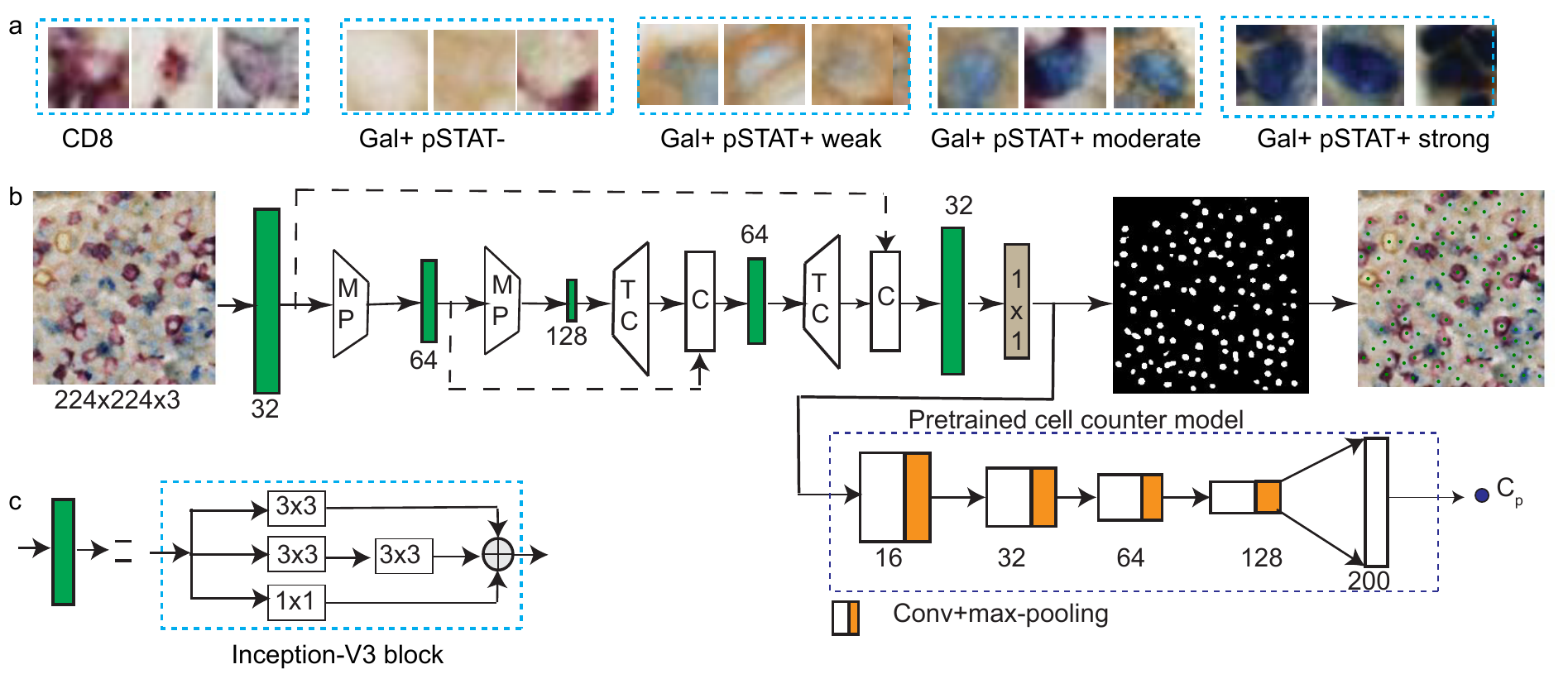}
	\end{minipage}
	\caption[]{a) Sample patches representing different types of cells. b) Schematics of ConCORDe-Net architecture. $3\  \text{x}\  3$ and $1 \text{x} 1$ indicate filter size of convolutional layers. TC = Transposed Convolution, MP = Max-pooling, C = Concatenate. The network has two outputs, probability map and predicted number of cells ($C_p$).  The probability map was thresholded using an empirically optimized threshold $T=0.85$ to convert to binary image. The center of every binary object represents center of a cell. c) Schematics of inception module.}
	\label{fig:detectionpipeline}
\end{figure}

Our proposed cell counter network is shown in Fig. \ref{fig:detectionpipeline}b.  It is a mapping function, $f:\mathbb{R}^{nxn} \rightarrow \mathbb{R}^{1}$, where \textit{n} is the size of the input patch, which is  $224$ in our case. It consists of feature extraction and regression parts. The feature extraction part is composed of four consecutive convolutional layers of $3\  \text{x}\  3$ filter size, and \textit{"same"} padding. The number of neurons in these layers are $\{16, 32, 64, 128\}$ respectively. Every convolutional layer was followed by max-pooling layer of size ($2 \ \text{x} \ 2$) with stride $2$ to reduce the dimensionality of features in the previous layer. The regressor part has a series of two dense layers of $\{200,\  1\}$ neurons. The output dense layer has one neuron which computes estimated number of cells in the input tensor or image. The activation of all convolutional and dense layers was set to \gls{relu}.

Parameters of all layers were randomly initialized using uniform glorot initialization \cite{Glorot}. Optimization of the parameters was done using Adam \cite{Kingma2014}, learning rate of $10^{-4}$. Initially, we have experimented with Euclidean loss \cite{Xue2016} and exponential loss functions. However, these suffer from loss explosion during the initial epochs and we came up with a new cell count loss ($C_l$) function in Equation (\ref{eq:cl}).
\begin{equation} \label{eq:cl}
	C_{l} = (1 - \frac{1}{1 + \frac{1}{B}\sum_{j=1}^{B}|C_{pj} - C_{tj}| })
\end{equation}
where the summation is over $B$ mini-batch images, $C_{pj}$ and $C_{tj}$ are predicted and true number of cells in  the $j^{th}$ image,  respectively. Fig. \ref{fig:roc_auc}a shows profile of $C_l$  as a function of cell count difference ($C_{p} - C_{t}$) and it is bounded between $0$ and $1$.

Before integrating the cell counter model to cell detection pipeline, it was trained and evaluated using pseudo-segmentation and number of cells as an input and output, respectively. To increase the amount of data, horizontal and vertical flipping were applied to all input training patches. The pseudo-segmentation is a binary image, however, when it is integrated with the cell detection model, a tensor of floating value will be fed. Thus, morphological and intensity deformation was applied as follows; Morphological erosion using rectangular structuring element of width $w=2$ was performed to every patch with a probability $p=0.4$, where \textit{p} and \textit{w} were empirically chosen.  Then, the images were multiplied by a random matrix of the same size as the image with an empirically chosen probability $p=0.4$. All elements in the random matrix were in range $[0.7, 1]$ to set pixel values between $0.7$ and $1$.
\subsection{Cell Detection} \label{sec:celldetect}
Fig. \ref{fig:detectionpipeline}b shows the proposed \gls{concoddenet} cell detection convolutional neural network. The input is $224 \text{x} 224 \text{x} 3$ size patch. The network has three parts; encoder, decoder and cell counter. The encoder-decoder section is extended version U-Net \cite{Ronneberger2015}. The standard U-Net architecture \cite{Ronneberger2015} uses VGG-style in its encoder and decoder section. We have proposed to use inception-v3 module shown in Fig. \ref{fig:detectionpipeline}c  instead of VGG block. The parallel and varying size  filters in inception block enables the network to extract multi-scale features in a given layer. The encoder contains three inception modules and the first two modules were followed by 2D max-pooling layers. The decoder is composed of transposed convolution, concatenation, and inception modules. The $1 \text{x} 1$ filter size convolutional layer at the end of the decoder is used to reduce the dimension of the tensor from $224 \text{x} 224 \text{x} 32$ to $224 \text{x} 224 \text{x} 1$. The output of the decoder was taken as cell location prediction map (P) and connected to the pretrained cell counter model (explained in Section \ref{sec:cellcount}), which generates predicted number of cells ($C_p$). Activation of all layers was set to \gls{relu}, but sigmoid for the last layer in the decoder section. Therefore, the cell detection architecture has two outputs, cell location prediction map and predicted number of cells.

The parameters of cell counter model were transfer learned from cell pseudo-segmentation as explained in Section \ref{sec:cellcount}. Parameters of the other layers were randomly initialized using uniform glorot initialization \cite{Glorot}, and optimized using Adam \cite{Kingma2014}, learning rate=$10^{-4}$ and an objective function shown in Equation (\ref{eq:dl}). Cell detection loss ($D_l$) in Equation (\ref{eq:dl}) has two parts. The first part is Dice overlap loss, and the second part is cell count loss.
\small
\begingroup\makeatletter\def\f@size{8}\check@mathfonts
\begin{equation}\label{eq:dl}
	D_{l}= (1- 2\frac{\sum_{j=1}^{B}\sum_{i=1}^{N}p_{ij}g_{ij}}{1  + \sum_{j=1}^{B}\sum_{i=1}^{N}p_{ij} + \sum_{j=1}^{B}\sum_{i=1}^{N}g_{ij}}) + K(1 - \frac{1}{1 + \frac{1}{B}\sum_{j=1}^{B}|C_{pj} - C_{tj}| })
\end{equation}
\endgroup
\normalsize
where summations in the first part is over batch size ($B$) images, and $N$ pixels of the ground truth image, $g_i\  \epsilon\  G$ and prediction map, $p_i \ \epsilon\  P$. The second part is same as Equation (\ref{eq:cl}), but weighted by empirically optimized constant $K=0.3$.

Horizontal and vertical flipping was applied to training patches to increase the amount and diversity of our data.   
\subsection{Cell Classification}
In our dataset, there were five types of cells: CD8, GAL8+ pSTAT-, GAL8+ pSTAT+ strong, GAL8+ pSTAT+ moderate, and GAL8+ pSTAT+ weak. GAL8+ pSTAT+ cells were divided based on the expression level of pSTAT into strong, moderate, and weak. However, discriminating among GAL8+ pSTAT+ cells is challenging, even for experts. Inspired by the principle of divide and conquer algorithm, we convert the problem into multi-stage classification. The first classifier (\textbf{classifier1}) differentiates between CD8, Gal8+ pSTAT-, and all GAL8+ pSTAT+ cells. Then, a second classifier (\textbf{classifier2}) was trained to further divide GAL8+ pSTAT+ cells in to GAL8+ pSTAT+ strong, GAL8+ pSTAT+ moderate, and GAL8+ pSTAT+ weak.

Both classifiers were trained using $28 \text{x} 28 \text{x} 3$ patches which can cover the whole cell area for the majority of the cells. Similar network architecture was used for both classifiers. The classifier has feature extraction and classification sections. The feature extraction part is a modified version of VGG architecture \cite{Simonyan2014} consisting of four convolutional layers of \{$32,\ 64\ 128\ 128$\} neurons with filters size $3\  \text{x}\  3$, stride $1$ and \textit{"same"} padding. Each convolutional layers were followed by $2\  \text{x}\  2$ max-pooling. The classification layer consisted of two dense layers of  \{$200,\ 3$\} neurons with dropout layer, rate=$0.3$ in between. Softmax activation was applied to the last dense layer and \gls{relu} for the other layers. Categorical cross-entropy objective function was applied. Uniform glorot \cite{Kingma2014} was applied to initialize parameters of the layers and optimized using Adam \cite{Kingma2014}, learning rate=$10^{-4}$. To handle class imbalance, in each mini-batch, an equal number of patches from all cell types were fed to the network and the number of iterations were determined by the number of patches in the most underestimated class. Moreover, runtime augmentation of flipping, and zooming with scale $s= [0.85\  1.15]$ was applied with a probability of $p=0.4$, where $s$ and $p$ were empirically optimized.
\section{Results and Discussion} \label{sec:result}
The proposed deep learning based unified cell detection and classification pipeline was evaluated on mIHC whole-tumor slide images. Implementation of the proposed approach was done in Python, and we used Keras API \cite{FrancoisCholletetal.2015} for development of the deep learning pipeline.

To investigate if \gls{cnn} can regress the number of cells from an input image, the proposed cell counter model was trained and then, evaluated on a test patches pseudo-segmentation image before integrating to ConCORDe-Net. Pearson correlation $r=0.999$ was obtained between the true and predicted number of cells. The high correlation supports that the proposed cell counter network can be used as a cell count approximation function.

Quantitatively, we evaluated ConCORDe-Net using standard metrics: precision, recall and F1-score. A detection was considered true positive if it lies with in an Euclidean distance of $8$ pixels ($2r$, where \textit{$r$} is in Equation(\ref{eq:annotation})) to a ground truth annotation. 

Moreover, we compared ConCORDe-Net with state of the art methods, MapDe \cite{Raza2018} and U-Net \cite{Ronneberger2015} as shown in Table \ref{tab:detect}. The same data augmentation as explained in Section \ref{sec:celldetect} was applied to all models depicted in the Table. U-Net \cite{Ronneberger2015} was trained to regress pseudo-segmentation explained in Section \ref{sec:annotation}. The output of \gls{cnn} models in Table \ref{tab:detect} is probability map that approximates pseudo-segmentation. The center of cells was regressed as follows from the probability map. Firstly, a global threshold maximizing F1-score was applied for each model to generate binary image. Secondly, hole filling morphological operation was applied to remove holes created after thresholding. Finally, the center of every connected component was computed which corresponds to center of a cell.
\begin{table}
	\centering
	\caption{Cell detection performance comparison. Model$1$ is a model after cell counter is removed from \gls{concoddenet}. U-Net \cite{Ronneberger2015} + Cell Counter is a \gls{cnn} after integrating cell counter \gls{cnn} to the original U-Net \cite{Ronneberger2015} architecture.}\label{tab:detect}
	\begin{tabular}{|l|l|c|c|}
		\hline
		Method    & Precision & Recall & F1-score\\
		\hline
		\textbf{\gls{concoddenet}}  &   0.854   &  \textbf{0.892}  &  \textbf{0.873} \\
		U-Net \cite{Ronneberger2015} + Cell Counter        &   0.872 	 &  0.837  & 0.854 \\
		Model$1$  &   \textbf{0.908}   &  0.80  &0.845 \\
		U-Net \cite{Ronneberger2015}        &   \textbf{0.908}   &  0.785  & 0.841 \\
		MapDe \cite{Raza2018}      &   0.804   &  0.876  & 0.838\\
		\hline
	\end{tabular}
\end{table}
ConCORDe-Net achieved the highest recall and F1-score compared to state of the art methods, MapDe \cite{Raza2018} and U-Net \cite{Ronneberger2015}. Moreover, in both \gls{concoddenet} and U-Net \cite{Ronneberger2015}, integrating cell counter \gls{cnn} has improved cell detection F1-score. For MapDe \cite{Raza2018}, we used the parameters that were specified in the paper and tuning the dimensions of “\textit{mapping filter}” might improve the result. 

Precision of ConCORDe-Net was lower than the three other methods due to the following reasons: 1) ConCORDe-Net identifies weakly stained cells that were missed by other methods, which could be missed by expert too. 2) Over-detection of large cells when there are more than one intensity peaks within the cell. We believe that these limitations could be improved by training and validating on a large cohort.
\begin{figure}
	\centering
	\begin{minipage}[b]{0.8\linewidth}
		\centering 
		\includegraphics[width=\textwidth]{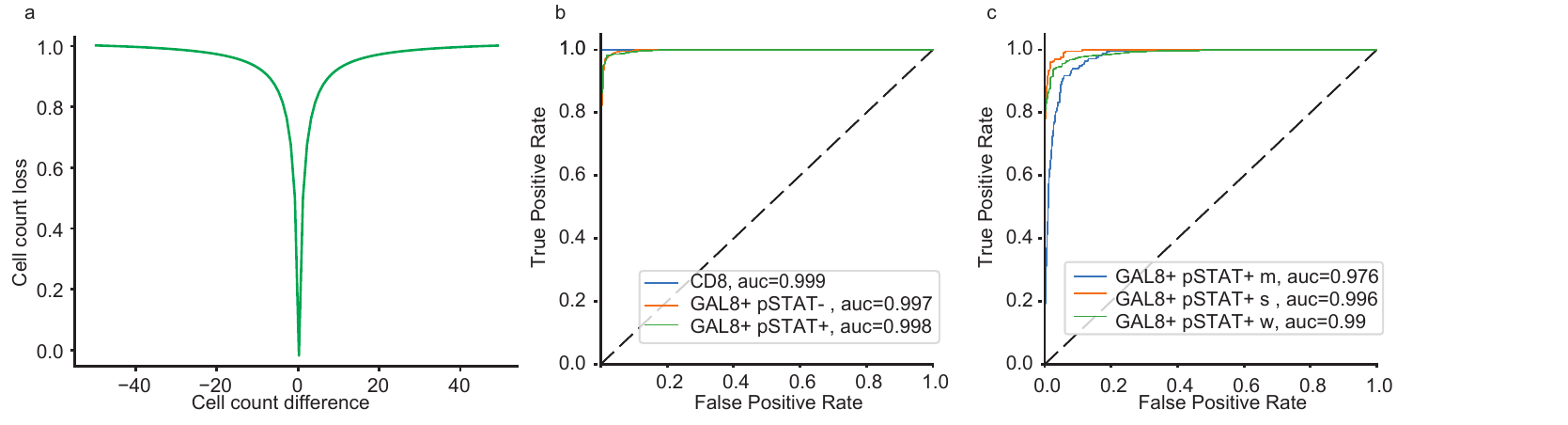}
		\centering
	\end{minipage}
	\caption[]{a) Cell count loss profile. ROC and AUC evaluation of b) classifier1 c) classifier2 on  test data. Where s=strong, m=moderate, w=weak}
	\label{fig:roc_auc}
\end{figure}

Performance of the proposed classifier models was qualitatively evaluated using \gls{roc}, \gls{auc}, accuracy, precision, recall, and F1-score on test data shown in Table \ref{tab:classificationdata}. \gls{roc} and \gls{auc} of \textbf{classifier1} are presented in Fig. \ref{fig:roc_auc}b. AUC value of greater than 0.99 was achieved for all cell types. Overall accuracy computed on the original distribution of data was found around $98\%$. Moreover, precision, recall and F1-score were all $0.98$. Fig. \ref{fig:roc_auc}c shows \gls{roc} and \gls{auc} of this \textbf{classifier2}. For all cell types, \gls{auc} value was higher than $0.97$ and overall accuracy of around $93\%$ was obtained. After cascading the two classifiers, overall accuracy of $96.5\%$  was achieved.
\begin{figure}
	\centering
	\begin{minipage}[b]{0.75\linewidth}
		% \centering
		\includegraphics[width=\textwidth]{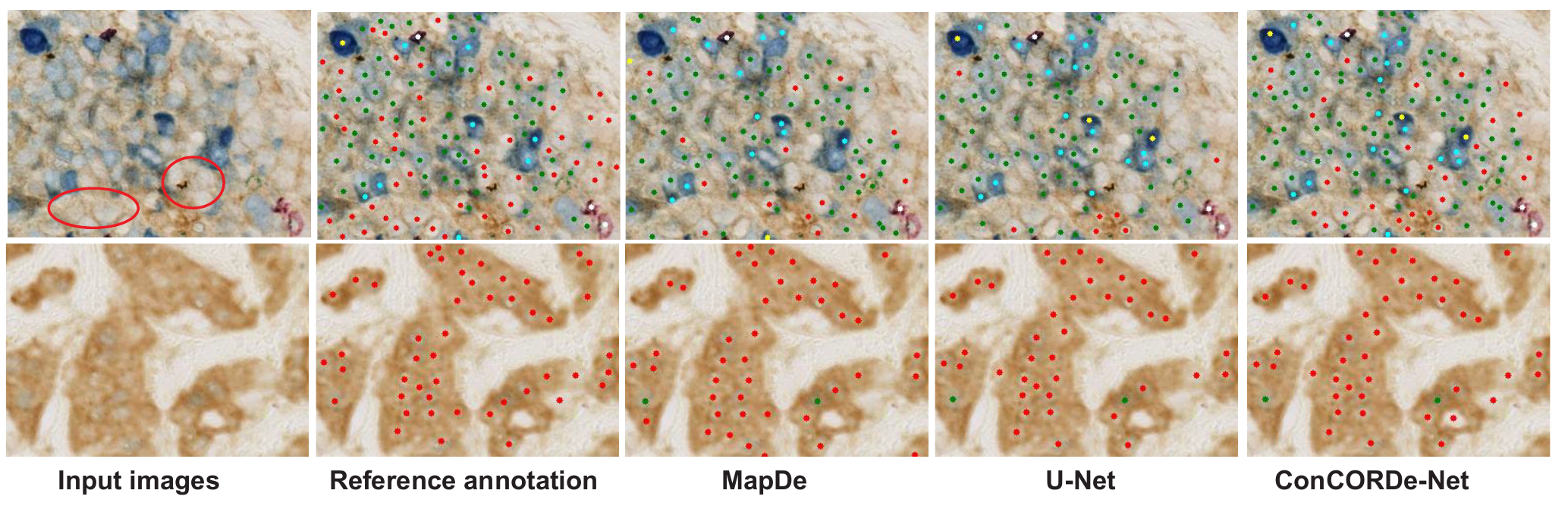}
	\end{minipage}
	%\vspace*{-3mm}
	\caption[]{Illustrative examples of the proposed unified cell detection and classification on test data, and comparison with state-of-the-art method, MapDe \cite{Raza2018} and U-Net \cite{Ronneberger2015}. White, red, yellow, cyan and dark green colored points represent CD8, GAL8+ pSTAT-, GAL8+ pSTAT+ strong, GAL8+ pSTAT+ moderate, and GAL8+ pSTAT+ weak cells, respectively. The red circles on the top left input images highlights cells that were missed by MapDe \cite{Raza2018} and U-Net \cite{Ronneberger2015}, but detected using \gls{concoddenet}.} 
	\label{fig:comparison}
\end{figure}

Fig. 3 shows a visual output of ConCORDe-Net followed by cell classification and comparison with MapDe \cite{Raza2018} and U-Net \cite{Ronneberger2015} which uses Dice overlap loss as an objective function. ConCORDe-Net is better in discerning touching cells with weak boundary gradient and weakly stained GAL8+ pSTAT- cells compared to MapDe \cite{Raza2018} and U-Net \cite{Ronneberger2015}. By regularizing the objective function with cell count, the network was able to learns patterns that can separate closely located cells and identify weakly stained cells.
\section{Conclusions}\label{sec:conclusions}
In this paper, we proposed a deep learning based unified cell detection and classification method in mIHC whole-tumor slide images of breast cancer. Cell count regularized CNN was employed for cell detection followed by multi-stage CNN to classify cells. The parameters in the cell detection architecture were learnt using a new objective function which optimizes dice overlap and cell count. F1 score of 0.873 was achieved on test data which outperformed state of the art methods MapDe \cite{Raza2018} and U-Net \cite{Ronneberger2015}. Our proposed approach is better in detecting closely located and weakly stained cells compared to MapDe \cite{Raza2018} and U-Net \cite{Ronneberger2015}. Moreover, $96.5\%$ classification accuracy was achieved. Our experiment shows that incorporating problem specific knowledge such as cell count improves robustness of the cell detection algorithm.
\section*{Acknowledgement}
This project was funded by the European Union's Horizon 2020 reaearch and innovation programme under the Marie Sklodowska-Curie grant agreement No 766030.

\end{document}